\documentclass[aps,superscriptaddress,floats,twocolumn,prb,showpacs,10pt]{revtex4-1}
\usepackage{graphicx}
\usepackage{amsmath}
\usepackage{color}

\begin{document}

\let\n=\nu
\let\o=\omega
\let\s=\sigma
\def\np{\n'}
\def\sp{\s'}
\def\EL{E_{L}}
\def\EN{E_N}
\def\ES{E_S}
\def\EM{E_M}
\def\ExN{\mbox{e}^{-\beta \EN}}
\def\ExM{\mbox{e}^{-\beta \EM}}
\def\ExL{\mbox{e}^{-\beta \EL}}
\def\ExS{\mbox{e}^{-\beta \ES}}
\def\c{(c_{\s})}
\def\cd{(c_{\s}^{\dagger})}
\def\cp{(c_{\sp})}
\def\cpd{(c_{\sp}^{\dagger})}
\newcommand{\bra}[1]{\langle #1|}
\newcommand{\ket}[1]{|#1\rangle}
\newcommand{\braket}[2]{\langle #1|#2\rangle}
\newcommand{\bvec}[1]{\ensuremath{\mathbf{#1}}}
\newcommand{\expval}[1]{\langle #1\rangle}
\renewcommand{\cal}[1]{\ensuremath{\mathcal{#1}}}
\newcommand{\dul}[1]{\ensuremath{\ushortdw{#1}}}
\newcommand{\sul}[1]{\ensuremath{\ushortw{#1}}}

\title{Atomic and itinerant effects at the transition metal x-ray absorption K-pre-edge exemplified in the case of V$_2$O$_3$}
\author{P. Hansmann}
\affiliation{Institute for Solid State Physics, Vienna University of Technology, 1040 Vienna, Austria}
\affiliation{Centre de Physique Th\'eorique, \'Ecole Polytechnique, CNRS-UMR7644, F-91128 Palaiseau, France}
\author{M. W. Haverkort}
\affiliation{Max Planck Institute for Solid State Research, 70569 Stuttgart, Germany}
\author{A. Toschi}
\affiliation{Institute for Solid State Physics, Vienna University of Technology, 1040 Vienna, Austria}
\author{G. Sangiovanni}
\affiliation{Institute for Solid State Physics, Vienna University of Technology, 1040 Vienna, Austria}
\author{F. Rodolakis}
\affiliation{Laboratoire de Physique des Solides, CNRS-UMR 8502, Universit\'e Paris-Sud, FR-91405 Orsay, France}
\affiliation{Synchrotron SOLEIL, L'Orme des Merisiers, Saint-Aubin, BP~48, 91192 Gif-sur-Yvette Cedex, France}
\affiliation{Material Science Division, Argonne National Laboratory, Argonne, Illinois 60439, USA}
\author{J. P. Rueff}
\affiliation{Synchrotron SOLEIL, L'Orme des Merisiers, Saint-Aubin, BP~48, 91192 Gif-sur-Yvette Cedex, France}
\affiliation{Laboratoire de Chimie Physique--Mati\`ere et Rayonnement, CNRS-UMR~7614, Universit\'e Pierre et Marie Curie, F-75005 Paris, France}
\author{M. Marsi}
\affiliation{Laboratoire de Physique des Solides, CNRS-UMR 8502, Universit\'e Paris-Sud, FR-91405 Orsay, France}
\author{K. Held}
\affiliation{Institute for Solid State Physics, Vienna University of Technology, 1040 Vienna, Austria}

\date{\today }

\begin{abstract}
X-ray absorption spectroscopy is a well established tool for obtaining information about orbital and spin degrees of freedom in transition metal- and rare earth-compounds. For this purpose usually the dipole transitions of the L- (2p to 3d) and M- (3d to 4f) edges are employed, whereas higher order transitions such as quadrupolar 1s to 3d in the K-edge are rarely studied in that respect. This is due to the fact that usually such quadrupolar transitions are overshadowed by dipole allowed 1s to 4p transitions and, hence, are visible only as minor features in the pre-edge region. Nonetheless, these features carry a lot of valuable information, similar to the dipole L-edge transition, which is not accessible in experiments under pressure due to the absorption of the diamond anvil pressure-cell. We recently performed a theoretical and experimental analysis of such a situation for the metal insulator transition of (V$_{(1-x)}$Cr$_x$)$_2$O$_3$. Since the importance of the orbital degrees of freedom in this transition is widely accepted, a thorough understanding of quadrupole transitions of the vanadium K-pre-edge provides crucial information about the underlying physics. Moreover, the lack of inversion symetry at the vanadium site leads to onsite mixing of vanadium 3d- and 4p- states and related quantum mechanical interferences between dipole and quadrupole transitions. Here we present a theoretical analysis of experimental high resolution x-ray absorption spectroscopy at the V pre-K edge measured in partial fluorescence yield mode for single crystals. We carried out density functional as well as configuration interaction calculations in order to capture effects coming from both, itinerant and atomic limits. 
\end{abstract}


\maketitle

\section{Introduction}
The process of X-ray absorption spectroscopy (XAS) is the excitation of a core--shell electron under specific selection rules. Experimental techniques evolved, together with the theoretical understanding and simulation capabilities over the last 25 years to maturity and have hitherto proven to be a valuable tool for probing the groundstate of transition metal (TM) and rare earth (RE) compounds. Advanced techniques make use of the polarization dependence of the absorption. Measuring the dependence of the absorption for circularly polarized light, called circular dichroism (CD), together with powerful sumrules provides a spectroscopic way of measuring the magnetic susceptibility \cite{thole92, Juhin10}. The dependence of the absorption for linearly polarized light, called linear dichroism (LD), on the other hand, has been successfully used to detect the orbital occupation in various systems such as high T$_c$ cuprates \cite{chen95}, Ce-based heavy fermion compounds \cite{hansmann08}, and also oxides of vanadium like VO$_2$ and V$_2$O$_3$ (Refs. \onlinecite{park00} and \onlinecite{haverkort05}).
 
For the transition metals the most informative excitations are obviously those involving 3d orbitals. In this respect, most of experimental and theoretical analysis have been done for the V$2p \rightarrow $V$3d$ transitions, the so called L--edges (at energies around $\sim350$ eV to $\sim950$ eV -- i.e. soft x--ray range). These transitions are dipole allowed and therefore have large cross-sections. Nonetheless, complementary valuable information can be extracted also from other core electron excitations of higher order like V$1s \rightarrow $V$3d$ quadrupole transitions located in the pre--edge region of the so called $K$-edge (at energies ranging from $\sim5$ keV to $\sim10$ keV for transition metals -- i.e. hard x--ray range). 

One of the remarkable advantages of XAS is the possibility of an element selective measurement\cite{deGroot_book}, since the absorption edges have distinct energies due to the element--specific binding energies of the core shell electrons. Important for the application of X--ray absorption as a spectroscopic technique is the development of models to simulate accurately the observed spectra: a quantitative analysis allows for the determination of various (near) ground state properties, including the valence, spin and orbital state of the atoms under investigation \cite{fink85}. These calculations are based on cluster models and have been successfully extended \cite{thole97,groot94,tanaka94} as to include the influence of the core--hole using the full atomic multiplet theory. The scheme relies on the localized nature of the exciton and consists basically in diagonalizing a configuration interaction (CI) Hamiltonian. The calculations are parameter--based and, hence, not \emph{ab initio}. Mostly, the parameters are fitted to experimental data in order to extract information. The ``cluster calculations'', as they are often called, became quite popular (especially for the L-- and M--edges) due to their -- sometimes outstanding -- ability to reproduce experimental data. \emph{Ab initio} methods, on the other hand, such as density functional theory in the local density approximation (LDA) often reproduce well the continuum part of the spectra or the main edge region of $K$-edges. However, the density functional methods fail when excitonic features become important. Hence, in the broadest sense the two methods are either tailored for the limit of strong (atomic) electronic correlations (CI) or the itinerant limit (LDA).

With this in mind the TM K-pre-edge region\cite{deGroot_book} is peculiar since it inherits a ``meeting point'' of the atomic- and itinerant limit which, by its own right, makes such a challenging analysis appealing. Yet, this is not the only motivation to obtain a better understanding of higher order transitions in pre-edge regions of TM K-edges. In fact, the main theoretical interest for the K-pre-edge transitions originates from the possibility of analyzing XAS experiments performed under external pressure: while the use of external pressure provides one of the cleanest ways to change the physical parameters of a compound, in such experiments the dipole allowed L-edge transition energies overlap with the energy scale of the absorbtion of the diamond pressure cell, so that the K-edge in the hard x--ray regime represents a technique which can provide information inaccessible with other methods\cite{lupinature}. This problem is a common one when phase diagrams/transitions are explored by means of applying pressure. And, naturally, it becomes a severe obstruction when orbital or spin degrees of freedom play a major role in the studied compound. A well known example of a pressure induced transition is the metal-insulator transition (MIT) in chromium doped vanadium sesquioxide (V$_{(1-x)}$Cr$_x$)$_2$O$_3$. \cite{mcwhan69, robinson75} First considered to be a textbook example for a simple single-band Mott insulator, it was later realized that the metal insulator transition upon doping pure V$_2$O$_3$ (metal to insulator) with chromium or applying pressure to the doped sample (insulator to metal) cannot be understood without incorporating crucial orbital physics\cite{castellani78A, castellani78B, park00, held01B, held01, tanusri09, poteryaev07, toschi10}. While this issue of orbital occupations was studied for the doping-induced transition a while ago employing the vanadium L-edge \cite{park00}, a thorough understanding of the pre-edge region in the vanadium K-edge is still missing, hampering the analysis of the pressure induced transition. In a recent work \cite{rodolakis10} we presented first experimental data for the vanadium K-edge as a function of temperature, doping, and prssure in a powder sample (for an extended experimental report see also Ref. \onlinecite{rodolakis11}). With a theoretical analysis by means of LDA+ dynamical mean field theory (DMFT) combined with CI cluster calculations we pointed out a clear experimental evidence showing the difference between the metal insulator transition upon doping or applying pressure\cite{rodolakis10, rodolakis09, rodolakis11}. In the present work we not only provide all details of the previous calculation, but we also perform a completely new analysis for LD-XAS of V$_2$O$_3$. Last but not least, we exploit the XAS analysis of this important example of a metal insulator transition in a multi-orbital system in order to provide a general description of the K-pre-edge characteristics, which are by no means restricted to the case of V$_2$O$_3$.

The paper is structured as follows: In section \ref{sec2} we recapitulate crystal and electronic structure of V$_2$O$_3$. In Section \ref{sec3} we give a pedagogical introduction to the theoretical background for the absorption process with emphasis given to the specifics of the K-pre-edge absorption. In the following section \ref{sec4} results of the density functional and configuration interaction are presented and compared to experimental data.

\begin{figure}[t]
  \begin{center}
  \includegraphics[width=0.45\textwidth]{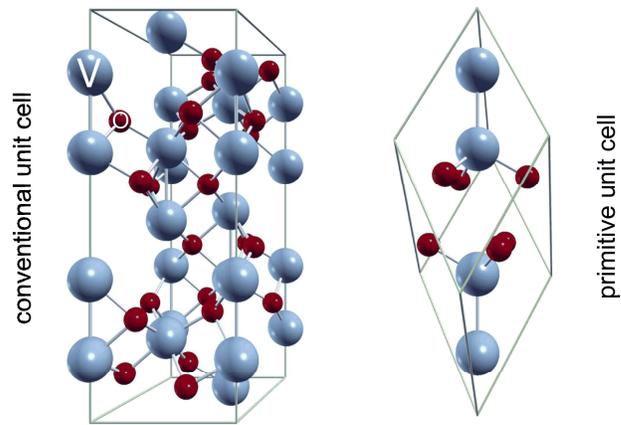}
  \end{center}
    \caption{(color online) Conventional (left hand side) and primitive (right hand side) unit cell of V$_2$O$_3$ in the corundum structure. The coordination polyhedron of the vanadium atom is an octahedron of oxygen ligands trigonally distorted along the C$^3$ axis (i.e. two opposing planes of the octahedron are ``squeezed'' together). The point group of a vanadium site is D$_{\rm 3d}$. Moreover, inversion symmetry along the closest V--V bond is broken which leads to an onsite mixing of V 3d and V 4p states.}
    \label{V2O3_crys}
\end{figure}

\begin{figure}[t]
  \begin{center}
  \includegraphics[width=0.45\textwidth]{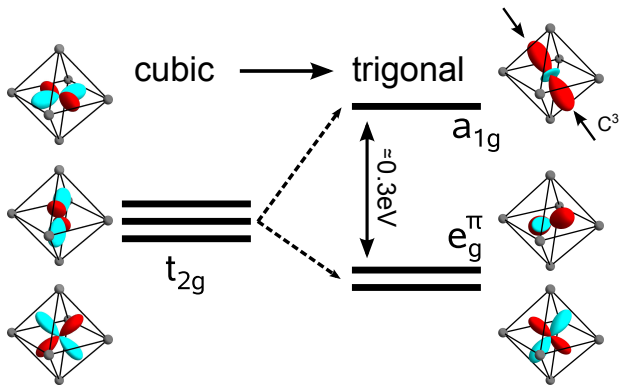}
  \end{center}
    \caption{(color online) Level splitting of the vanadium $t_{2g}$--states: due to the trigonal distortion (sketched for the $a_{1g}$ case with the arrows as a compression along the threefold C$^3$-axis of the octahedron) the $t_{2g}$--states split up in an $e^\pi_g$ doublet and an $a_{1g}$ singlet. From LDA results we estimate approximately 0.3 eV for this splitting. The plotted orbitals are spherical harmonic functions which display the symmetry of the states.} 
    \label{V2O3_lvl}
\end{figure}

\section{Crystal and electronic structure of V$_2$O$_3$}
\label{sec2}
At ambient conditions V$_2$O$_3$ is metallic and crystallizes in the corundum structure with four vanadium atoms in the primitive unit cell. The conventional and the primitive unit cells are sketched in Fig.~\ref{V2O3_crys} where it can be seen, that the vanadium atoms form ``pairs'' oriented along the crystallographic c--axis. Upon cooling below 150K, antiferromagnetic order sets in and the system becomes insulating; the onset of the magnetic order is accompanied by a transition from corundum to monoclinic structure. In addition, the system can be driven through (or further away from) the Mott MIT by doping with chromium or titanium or the application of external pressure. Above the N\'{e}el temperature the crystal structure does not change symmetry as a function of pressure or doping. However, upon Cr doping a first order isostructural metal--to--insulator (PM--to--PI) transition takes place (terminating in a critical point arround $\approx 400$K) which evoked several theoretical attempts to describe this MIT in the spirit of a single band Hubbard model. While the MIT is associated to changes in the lattice structure and the atomic positions \cite{mcwhan69,robinson75}, it is important to notice that X-ray diffraction showed that for a given temperature the structure within one phase \emph{does not change upon doping} \cite{robinson75} (but only with T \cite{robinson75, baldassarre08} - for complementary information from photoemission see Ref. \onlinecite{rodolakis09PRL}). It was later also observed by Park \emph{et al.} with vanadium $L$-edge x-ray absorption spectroscopy that this holds also for the electronic ground state of the system (see Table 1 of Ref. \onlinecite{park00}). Often it is assumed, that doping and pressure can be seen as equivalent routes through the phase diagram. Recently, however, experimental evidence and theoretical analysis suggest to discard the common wisdom of the pressure/doping equivalence\cite{rodolakis10}.

The electronic configuration of atomic vanadium is [Ar]3d$^3$4s$^2$, which means, that in the three--valent oxidation state one finds a 3d$^2$ configuration realized. In the corundum type structure the vanadium atoms are coordinated by oxygen ligands in a trigonally distorted octahedral fashion (left hand side of Fig.~\ref{V2O3_crys}). Moreover, the crystal field breaks one significant point symmetry on the vanadium sites, namely inversion in the c-direction. This is related to the aforementioned V-pairs resulting in different distances of the neighboring vanadium atoms along the c-axis. As a result we find a slight \emph{onsite} mixing of vanadium 3d- and 4p-states. While this effect is generally negligible for a discussion of the ground state composition, it will be of great importance for the selection rules of the polarization dependent XAS results which we discuss in the following sections.

The major part of the crystal field is of cubic symmetry and, hence, splits the d--states into the lower lying t$_{2g}$ and the higher lying e$_g$ states. The trigonal distortion acts like a compression along one of the three--fold axes of the octahedron (i.e. squeezing two opposite sides together). As a result the degeneracy of the lower lying $t_{2g}$ states is lifted and they are split into a single $a_{1g}$ and an $e_{g}$ doublet. In order to distinguish the two $e_g$ doublets, each of them gets an additional index indicating the type of bonds (either $\sigma$ or $\pi$) they form with the ligands: the high lying ''cubic'' $e_g$ (not split by the trigonal distortion) gets an additional ''$\sigma$'' and reads from now on $e_g^\sigma$ whereas the lower one gets an additional $\pi$ and reads from now on $e_g^\pi$. The level splitting, together with a plot of the respective angular part of the (atomic) wave function, is sketched in Fig.~\ref{V2O3_lvl}. 
Since the $e_g^\sigma$ are higher in energy due to the crystal field, the two vanadium d--electrons populate the t$_{2g}$ levels. One of the crucial aspects concerning the understanding of the MIT in the paramagnetic regime is the specific occupation of these t$_{2g}$ states. Experimental evidence suggests\cite{moon70, paolasini99, dimateo02, park00} that the ground state of the system should be described as a $S=1$ state consisting of a mixture of $a_{1g}$ and $e_{g}^\pi$.
Moreover, it is precisely the coefficients in the linear combination of $a_{1g}$ and $e_{g}^\pi$ for the ground state which allow for a quantitative distinction of the PM, PI, and AF phases. The XAS vanadium $L$--edge study of Park \emph{et al.} explored the phase diagram by means of temperature and doping and summarized the respective ratios of $a_{1g}$ and $e_{g}^\pi$ in table 1 of Ref. \onlinecite{park00}. As we mentioned earlier, their results turned out to be consistent with the X-ray diffraction data for the lattice of Robinson \cite{robinson75}, and showed that, within the PM- and within the PI - phase, there is no change in the ground state composition for different doping\cite{rodolakis10}.

\section{X-ray absorption and dichroism for the vanadium K-edge}
\label{sec3}
In our recent work\cite{rodolakis10} we identified a robust ground state probe by calculating the ratio of the first two peaks in the K-pre-edge region. In Ref.~\onlinecite{rodolakis10} we discussed the origin of the ratio change in the case of a powder sample. However, a huge amount of extra information can be obtained by incorporating dichroic effects in our considerations. In fact, we will not only see that the dichroism of the K-pre-edge XAS fully supports the claim made for the powder spectra, but our theoretical results explicitely predict non-trivial interference between dipole and quadrupole transitions. The origin of such interference effects can be found in the earlier mentioned onsite mixing of vanadium d- and p-states: although the amount of p-character of the vanadium ''d-orbitals'' is small, the dipole allowed transition of the 1s core electron into this p-part makes the effect clearly visible in the K-pre-edge.  However, the interference effect is not directly accessible in standard XAS since for these an average over all vanadium atoms in the unit cell is taken, leading to a symmetry conditioned cancellation. In fact, $\bvec{k}$-dependent absorption techniques could access the predicted intereference effects.\\

Let us start by recalling the basic physics behind linear dichroism in X-ray absorption. The reason for linear dichroism and its relation to the occupancy of crystal field eigenstates can be understood intuitively even in a one--electron picture. Consider the following example: We want to promote a core $s$--electron via a dipole transition to a valence state of $p_z$--symmetry\cite{stohr_book}. Following Fermi's golden rule in the sudden approximation, the spectrum of such a process is proportional to the square of the integral $\bra{s}D_{x,y,z}\ket{p_z}$ with dipole operator $D_{x,y,z}\propto(x,y,z)$, $\bra{s}\propto 1$ and $\ket{p_z}\propto z$ in Cartesian coordinates. If the integrand is an odd function the integral vanishes. Hence, the \emph{only} dipole transition operator for which the integral remains finite is the $D_z\propto z$. The same holds, of course, also for the corresponding transitions to $p_x$-- or $p_y$--states, i.e., with $x$ and $y$ polarized light one can excite into the $p_x$-- or $p_y$--states respectively. Now, let the ground state of our hypothetical system be a single occupied $p_z$--orbital and empty $p_x$ and $p_y$ orbitals. Hence, adding an electron to the $p_z$--orbital will cost additional coulombic energy, so that the peak of the $z$--polarized spectrum (corresponding to the $D_z$ transition operator) lies at higher energies compared to the $D_{x/y}$ spectra showing also only half the intensity due to its occupancy. This means valuable information about the ground state of the system is encoded in the LD spectra. In other words: The polarization of the x-ray light adds \emph{more selection rules} to the absorption process which can be used for a more detailed study of the ground state properties.

\begin{figure*}
  \begin{center}
    \includegraphics[width=0.8\textwidth]{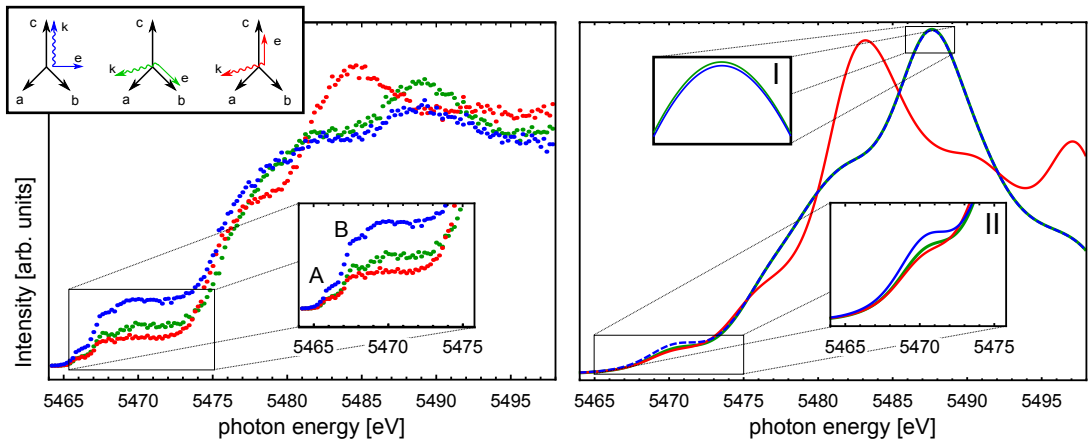}
  \end{center}
  \caption{(color online) Polarization dependent experimental data for the vanadium $K$-edge (left) compared to calculated LDA data (right). The color of the respective plot codes the geometry of the measurement, i.e., the transition operator, explained in the sketch on the left hand side. We can see the effects of the inversion symmetry breaking along the V--V bonds clearly in the insets, where we observe the splitting of spectra which should be equivalent by definition in a system without the symmetry breaking (see Eq. \eqref{symCond}). Note that the differences in the intensities of the largest peaks can be attributed to self absorption effects, which are not included in the calculation.} \label{V2O3_ldavsmainedge}
\end{figure*}

Let us summerize now the most important transitions for the V $K$--edge: The main edge consists mostly of dipole transitions of the s--core electron into the unbound V 4p--states, whereas transitions to the localized and correlated 3d--states via higher order processes contribute to the pre--edge. We formally write the absorption process employing Fermi's golden rule as:
\begin{equation}\label{transproc}
\left|\sum_{\Psi_{\text{final}}}\bra{\Psi_{\text{final}}}T_{\boldsymbol\epsilon}\ket{\Psi_{\text{initial}}}\right |^2\delta(\omega-E_{\rm initial}+E_{\rm final}).
\end{equation}

This quantity is directly related to the spectral intensity. The transition operator $T_{\boldsymbol\epsilon}$ describes the coupling of the photon to to our system and can generally be written as \cite{loudon83,mercouris97}:
\begin{equation}\label{transop}
 T_{\mathbf{\epsilon}}=e^{i\bvec{k}\cdot \bvec{r}} \bvec{p}\cdot {\boldsymbol\epsilon},
\end{equation}
where $\bvec{k}$ and ${\boldsymbol\epsilon}$ are the light propagation and polarization vectors and $\bvec{r}$ and $\bvec{p}$ are the electron position and momentum operators. Equation \eqref{transop} can be understood in terms of different multipole transitions. Expanding $e^{i\bvec{k}\cdot \bvec{r}}$ in terms of multipoles gives:
\begin{equation}
e^{i\bvec{k}\cdot \bvec{r}} = \sum_{l=0}^{\infty}\sum_{m=-l}^{l} \imath^l (2l+1) j_l(k r) C_{m}^{(l)^*}(\theta_k,\phi_k)C_{m}^{(l)}(\theta_r,\phi_r),
\end{equation}
with $j_l$ being a spherical Bessel function of order $l$; $k$, $\theta_k$, $\phi_k$, and $r$, $\theta_r$, $\phi_r$ the length and angular variables of the corresponding vectors $\mathbf{k}$, and $\mathbf{r}$, and finally $C_m^{(l)}$ a renormalized spherical harmonics $C_m^{(l)}=\sqrt{4 \pi/(2 l + 1)}Y_m^{(l)}$. If one furthermore replaces the momentum operator $\mathbf{p}$ by the position operator $\mathbf{r}$, using the commutation relations $\bvec{p}=\imath m/\hbar[H,\bvec{r}]$ one finds for the transition operator:
\begin{align}
T_{{\boldsymbol\epsilon},\mathbf{k}} =& j_0(k r) ( \mathbf{r} \cdot {\boldsymbol\epsilon} ) \\
\nonumber & + 3 \imath j_1(k r) (\mathbf{r} \cdot {\boldsymbol\epsilon}) (\mathbf{r} \cdot \mathbf{k}) + \text{higher order terms}\\             
\end{align}
with $\mathbf{r}$, $\mathbf{k}$, and $\boldsymbol\epsilon$ denoting unit vectors in the direction of the position, momentum, and light polarization, respectively. 

In order to rewrite the full transition operator $T_{{\boldsymbol\epsilon},\mathbf{k}}$ in terms of cubic-harmonics, we first define the multipole vector $M$ consisting of the Dipole and Quadrupole operators: 

\begin{equation}
M=(D_x,D_y,D_z,Q_{yz},Q_{xz},Q_{xy},Q_{x^2-y^2},Q_{z^2}) \label{Mvec}
\end{equation}.

The dipole operators can be expressed as:
\begin{eqnarray}\label{dipole}
\nonumber  D_x&=& j_0(k r) x \\
\nonumber  D_y&=& j_0(k r) y \\
           D_z&=& j_0(k r) z.
\end{eqnarray}
The quadrupole operators read:
\begin{eqnarray}\label{quadrupole}
\nonumber Q_{yz}&=& \imath 6 j_1(k r) yz\\
\nonumber Q_{xz}&=& \imath 6 j_1(k r) xz\\
\nonumber Q_{xy}&=& \imath 6 j_1(k r) xy\\
\nonumber Q_{x^2-y^2}&=&\imath 6 j_1(k r) \sqrt{1/4}(x^2-y^2)\\
          Q_{z^2}&=&\imath 6 j_1(k r) \sqrt{3/4}(z^2-1/3).
\end{eqnarray}

In this basis we can define a generalized polarization vector taking into account only dipole and quadrupole terms, i.e., neglecting higher order terms, by means of an eight-component vector:
\begin{eqnarray}
\varepsilon&=&\big( \epsilon_x,\quad \epsilon_y,\quad \epsilon_z,\quad  (1/2)(\mathbf{\epsilon}_y\mathbf{k}_z+\mathbf{\epsilon}_z\mathbf{k}_y),\\
\nonumber   &&\quad (1/2)(\mathbf{\epsilon}_x\mathbf{k}_z+\mathbf{\epsilon}_z\mathbf{k}_x),\quad (1/2)(\mathbf{\epsilon}_x\mathbf{k}_y+\mathbf{\epsilon}_y\mathbf{k}_x),\\
\nonumber &&\quad \sqrt{1/4}(\mathbf{\epsilon}_x\mathbf{k}_x-\mathbf{\epsilon}_y\mathbf{k}_y),\quad \sqrt{3/4}(\mathbf{\epsilon}_z\mathbf{k}_z-(1/3)\boldsymbol\epsilon\cdot\mathbf{k})\big).
\label{EPSvec}
\end{eqnarray}
The full transition operator $T_{\epsilon,\mathbf{k}}=M \cdot\varepsilon$ is then given by a dot product of \eqref{Mvec} and \eqref{EPSvec}. Finally, the conductivity tensor can be computed as:
\begin{eqnarray}\label{sigma}
\nonumber\sigma_{i,j} &=& \sum_{\Psi_{\text{final}}}\bra{\Psi_{\text{initial}}}M_{j}^{*}\ket{\Psi_{\text{final}}}\bra{\Psi_{\text{final}}}M_{i}\ket{\Psi_{\text{initial}}}\\
&&\times\delta(\omega-E_{\rm initial}+E_{\rm final}).
\end{eqnarray}
Knowing the conductivity tensor, one can straight forwardly calculate an absorption spectrum as $-\mathrm{Im}[\varepsilon^T\cdot\sigma\cdot\varepsilon]$. 

\subsection*{Linear dichroism at the vanadium $K$--edge}
Let us now turn to the linear dichroism of the vanadium $K$--edge. In Fig.~\ref{V2O3_ldavsmainedge} on the left hand side we show the experimental spectra (dots) of the V $K$--edge for three different polarizations of the x--ray light. In the upper inset we sketch the polarization of each spectrum with respect to the trigonal (local vanadium) reference frame $(a,b,c)$ ($\hat{\bvec{a}}$ and $\hat{\bvec{b}}$ form an angle of 120 degree, point towards the neighbouring V atoms in the plane and are both perpendicular to $\hat{\bvec{c}}$ which points in the direction of the out of plane V pairs). Since we would like to express the transition operators for each polarization in terms of the dipole and quadrupole operators \eqref{dipole} \& \eqref{quadrupole}, we translate to Cartesian coordinates:
\begin{align}
a=&x & c=&z\\
b=&\frac{\sqrt{3}}{2}y-\frac{1}{2}x & y=&\frac{1}{\sqrt{3}}(a+2b)
\end{align}
Now we can write down the dipole and quadrupole operators for each orientation in terms of \eqref{dipole} and \eqref{quadrupole}
\begin{align}
D_{\text{\textcolor{blue}{b}}}=&D_x&Q_{\text{\textcolor{blue}{b}}}=&Q_{yz}\\
D_{\text{\textcolor{green}{g}}}=&\frac{\sqrt{3}}{2}D_y+\frac{1}{2}D_x&Q_{\text{\textcolor{green}{g}}}\equiv&Q_{x^2-y^2}\\
D_{\text{\textcolor{red}{r}}}=&D_z& Q_{\text{\textcolor{red}{{r}}}}=&\frac{\sqrt{3}}{2}Q_{xz}+\frac{1}{2}Q_{yz}
\end{align}
Employing the crystal symmetry $a\leftrightarrow b$ we find straight forward
\begin{align}
D_{\text{\textcolor{blue}{b}}}=&D_{\text{\textcolor{green}{g}}} &Q_{\text{\textcolor{blue}{b}}}=&Q_{\text{\textcolor{red}{r}}}
\label{symCond}
\end{align}
Evidently, while for dipole transitions we find the blue (b) and green (g) polarization to be equivalent, in quadrupole transitions blue (b) and red (r) polarization should be indistinguishable.

However, when we inspect the experimental spectra closely we find these symmetries obviously violated: The lower inset of Fig.~\ref{V2O3_ldavsmainedge} (left hand side) shows a zoom of the pre--edge region where clearly all three spectra are different. Moreover, even in the main edge we find the ``blue/green'' dipole symmetry violated. As already mentioned in section \ref{sec2} the reason for this violation can be understood intuitively with the sketch of the primitive unit cell (Fig.~\ref{V2O3_crys} right hand side). Along the connection line of two vanadium atoms in the primitive unit cell, i.e., parallel to the crystallographic c--axis the vanadium site lacks inversion symmetry, since the distances to the ``upper'' and ``lower'' V--neighbors is different. The consequence of breaking such a fundamental symmetry can be quite dramatic, as we will see. Namely, the V 3d--states are no longer pure eigenstates of the system and can mix with V 4p--states \emph{onsite}. This means, that the formerly pure d--states get a tiny dipole moment and the quadrupole/dipole selection rule no longer holds strictly. This mixing will explain why the pre--edge is not entirely quadrupole in character and the main--edge not entirely dipole. Let us remark that this mixing is, in fact, small enough to be negligible for the formulation of the groundstate in terms of  $a_{1g}$ and $e_{g}^\pi$ occupation. In the absorption experiment, however, the smallness of the mixing is compensated by the difference in the matrix-element for dipole and quadrupole transitions. As a consequence, even a tiny amount of p-character is visible in the pre-edge, because it is dipole allowed: 

\begin{figure*}
  \begin{center}
    \includegraphics[width=\textwidth]{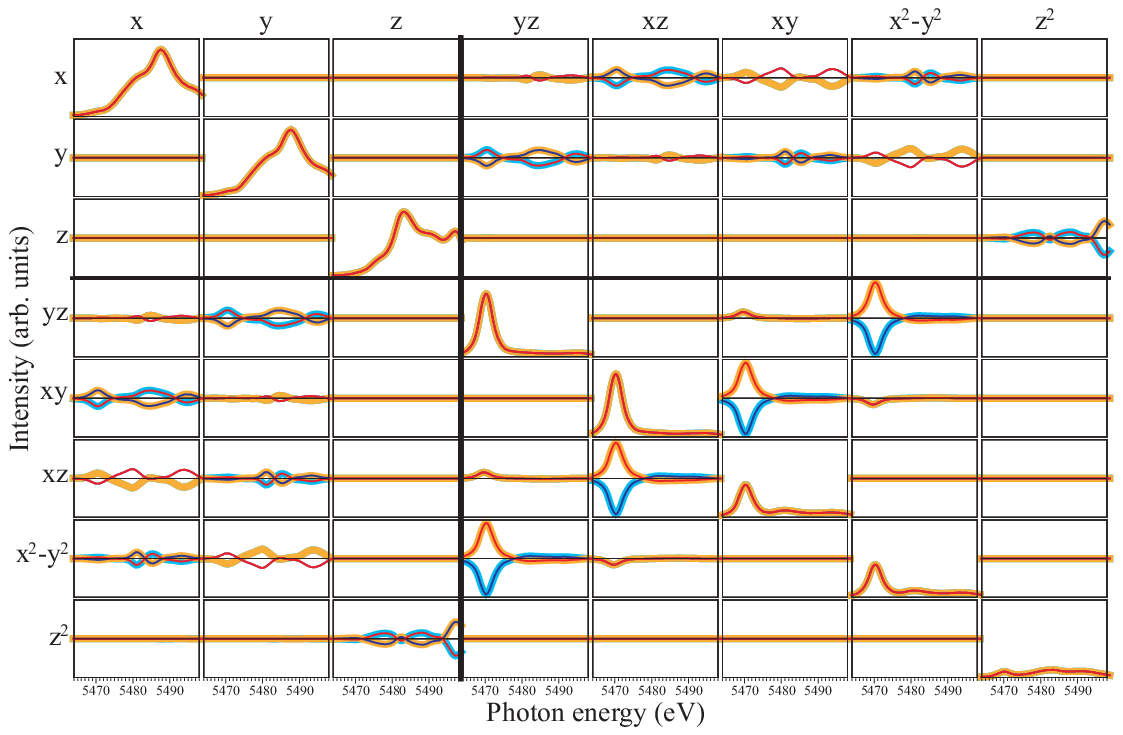}
  \end{center}
  \caption{(color online) Conductivity tensor $-\mathrm{Im}[\sigma]$ (see Eq.\eqref{sigma}) of V$_{2}$O$_{3}$ for dipole and quadrupole transitions at x-ray $K$-edge energies. Each panel shows a spectral function (for each of the four V atoms in the unit cell), representing an element of the conductivity tensor. The measured spectrum for a given experimental geometry is a linear combination of the tensor elements given by the generalized polarization (Eq. \eqref{EPSvec}) and the conductivity tensor as $-\mathrm{Im}[\mathbf{\varepsilon}^* \cdot \sigma \cdot \mathbf{\varepsilon}]$.} \label{V2O3_SigmaArray}
\end{figure*}

\begin{equation}\label{transproc2}
  \begin{split}
 &\sum_{\Psi_{\text{f}}}\left|\bra{\Psi_{\text{i}}}\hat{P}\ket{\Psi_{\text{f}}}\right|^2\delta(\omega-\Delta E)\\
 &=   \sum_{\Psi_{\text{f}}}\left|\bra{\Psi^{\rm core}_{1s}}(D+Q)(a\ket{\Psi_{4p}}+b\ket{\Psi_{3d}})\right|^2\delta(\omega-\Delta E)\\
 &=   \sum_{\Psi_{\text{f}}}\left|\bra{\Psi^{\rm core}_{1s}}(a D\ket{\Psi_{4p}}+b Q\ket{\Psi_{3d}})\right|^2\delta(\omega-\Delta E)
 \end{split}
\end{equation}
where we defined $\Delta E=E_f-E_i$ as the energy difference of initial and final state.
Now the spectral intensity of \eqref{transproc2} involves a dipole spectrum $a^2\left|\bra{\Psi^{\rm core}_{1s}}D\ket{\Psi_{4p}}\right|^2$ and a quadrupole spectrum $b^2\left|\bra{\Psi^{\rm core}_{1s}}Q\ket{\Psi_{3d}}\right|^2$. However, due to the mixed character of the final state $\ket{\Psi_{4p}}+\ket{\Psi_{3d}}$, both parts are entangled. A quantum interference term of the form
\begin{equation}\label{interference}
ab\bra{\Psi_{3d}}Q^*\ket{\Psi^{\rm core}_{1s}}\cdot\bra{\Psi^{\rm core}_{1s}}D\ket{\Psi_{4p}}
\end{equation}
emerges which has to be taken into account explicitly. In Fig.~\ref{V2O3_SigmaArray} we show a matrix representation of the spectra resolved in diagonal elements, i.e. \emph{pure} dipole (rows/columns 1-3) and quadrupole (rows/columns 4-8) transitions and the off-diagonal interference terms. The different colors in each of the panels encodes the index of the four vanadium atoms in the unit cell. It is important to notice that when an average over all atoms is taken, as it is done in the present experiment, the interference terms cancel out exactly and are not directly accessible anymore. This is related to the fact that although inversion symmetry is broken for the pointgroup of a single vanadium site, it is not broken for the unit cell of the corrundum structure. Note that this would be different in the case of x-ray \emph{diffraction} in which one measueres not the sum of $-Im[\epsilon \cdot \sigma \epsilon]$ over all sites, but $\sum e^{\imath k \cdot r_i} F_i$. This means a phase dependent sum over the site depenent scattering tensor, whereby $F=\omega \sigma$. In this case the inteference terms will not cancel out due to summation\cite{Tsvetkov04}.

The main edge, i.e., where excitonic features and the many--body coupling to the core hole are negligible, can be well described within the LDA. Since the \emph{ab initio} LDA calculation incorporates the symmetry breaking in the structural input, the information about the resulting d--p mixing is already \emph{included} in the eigenstates of the LDA Hamiltonian.

In order to calculate the XAS spectrum we have to consider only the muffin tin sphere around the vanadium atom $R_{\rm MT}$ since the V 1s core hole wave function is zero for $r>R_{\rm MT}$. We calculate projections of the LDA wave functions on the V 3d and V 4p subspace within the sphere:
\begin{equation}\label{LDAfun}
\Psi^{\rm LDA}_{l,m}=\cal{D}_{l,m}(\omega)R_{lm}(r)Y_{l}^{m}
\end{equation}
where we neglected the energy dependence of the radial part of the wave functions, and the energy dependent $\cal{D}_{l,m}(\omega)$ stems from the projection. The square of $\cal{D}_{l,m}(\omega)$ is in fact the so-called \emph{partial density of states}. In terms of \eqref{LDAfun} the integrals of \eqref{transproc2} separate into three parts: i) partial DOS $\cal{D}^2_{3d(4p)}(\omega)$ and ``interference terms''$\cal{D}_{3d}(\omega)\cdot\cal{D}_{4p}(\omega)$ ii) Radial transition probabilities of the form
\begin{equation}
\int_0^{R_{\rm MT}} r^2dr \cdot r R^*_{1s}(r)R_{4p}(r)
\end{equation}
for the dipole part, and
\begin{equation}
\int_0^{R_{\rm MT}} r^2dr \cdot r^2 R^*_{1s}(r)R_{3d}(r)
\end{equation}
for the quadrupole part. And finally iii) angular integrals like
\begin{equation}
\int d\Omega \; Y_{l}^{m} O_{l'}^{m'} Y_{l''}^{m''}
\end{equation}
where $O_{l'}^{m'}$ is the angular part of the specific transition operator. The angular integrals yield the selection rules for the transition.

\section{LDA and CI results}
\label{sec4}

In Fig.~\ref{V2O3_ldavsmainedge} on the right hand side, we present LDA spectra for the each transition symmetry.  The overall agreement for the main edge is very satisfying. Especially effects of the 3d/4p mixing are captured well by the calculation and yield interesting features. As a signature of this mixing we find a splitting of the three spectra (see insets) in the experimental and calculated data, which would not be present in pure dipole/quadrupole transitions.

In the pre--edge region the LDA calculation misses, obviously, the excitonic features (the first two peaks) which involve many-body states beyond LDA. Nonetheless, the LDA captures well the correct trends in the pre--edge region around 5470eV where transitions to the $e_g^\sigma$, the least correlated and most itinerant among 3d-states, are located. In order to use the ``rule of thumb'' from Ref.~\onlinecite{rodolakis10}, i.e., the ratio of the first two excitonic peaks (A and B) as probe of the $\alpha\ket{e_g^{\pi},e_g^{\pi}}+\beta\ket{e_g^{\pi},a_{1g}}$ ($\alpha^2+\beta^2=1$) ground state, we have to include the 3d--4p mixing in our cluster calculation.
This can be done by taking this mixing, i.e. the overlap integrals of V 3d and V4p, as a fitting parameter into account. The proper \emph{symmetry} (but not the quantitative values!) of the potential we obtain from a Madelung calculation. As a technicality we remark that the $t_{2g}$ and $e_g^\sigma$ spectra were calculated separately and summed up later. This was done in order to account for the big differences we expect for the broadening of the excitations due to the much stronger non--local hybridization of the $e_g^\sigma$ orbitals with the oxygen ligands. Moreover, we calculated the spectra for all four vanadium atoms in the unit cell and averaged them. While locally the vanadium atoms are equivalent, they differ with respect to an external frame of reference given by the propagation and polarization vector of the x-ray light.

\begin{figure}
  \begin{center}
    \includegraphics[width=0.45\textwidth]{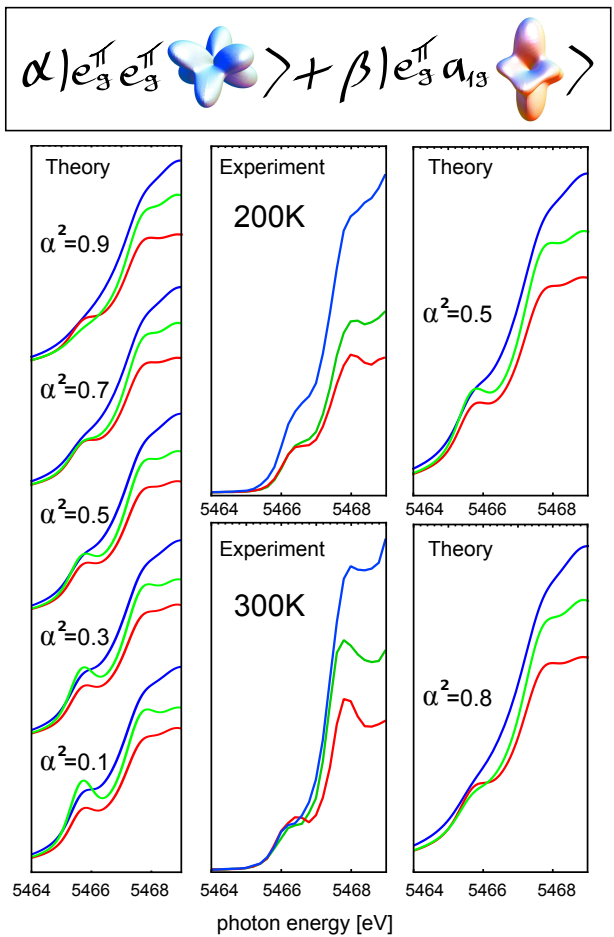}
  \end{center}
  \caption{(color online) Polarization dependent spectra in the pre--edge region of the vanadium $K$-edge. In the panel on the left hand side we show a ``map'' of theoretically calculated full multiplet spectra for different compositions of the ground state $\alpha\ket{e_g^{\pi},e_g^\pi}+\beta\ket{e_g^{\pi},a_{1g}}$ (with $\alpha^2+\beta^2=1$). Employing this map we find the best agreement with the experimental data (middle panel) for values of $\alpha^2=0.5$ for the PM phase (200K) and $\alpha^2=0.8$ for the PI phase (300K). This provides a further confirmation that the LDA+DMFT densities\cite{rodolakis10} are a good starting point for the CI calculation. \label{V2O3_mixtable}}
\end{figure}

In Fig.~\ref{V2O3_mixtable} we report the results of our CI together with experimental data for the PM and PI phase. On the left hand side of the figure we show a ''map'' of simulated spectra for different values of $\alpha^2$. Overall we observe that we can capture the trend of the spectrum hierarchy also in the cluster calculation quite well. This time, however, the \emph{itinerant} states are beyond the basis set of our CI calculation of a small cluster and we cannot expect quantitative agreement for energies in the region of the $e_g^\sigma$ or higher. On the other hand, we now capture the excitonic peaks A and B correctly which we need as a ground state probe. Let us remark at this point, that the $e_g^\sigma$--states can be captured from both sides, LDA and CI, only qualitatively. In fact, they can be understood to represent a kind of ''intermediate'' regime -- neither truly localized nor totally itinerant. The panel on the right hand side shows the color coding for the three polarization directions.

In the two middle panels of Fig.~\ref{V2O3_mixtable} we show the experimental spectra compared to the theoretical CI results which we obtained with the same mixing parameters we used for the isotropic spectra, i.e., very close to our previous DMFT results\cite{rodolakis10}. The agreement between experiment and CI calculations is quite satisfying. Moreover, we observe that the change of the ratio between peak A and B, i.e., our most important probe, going from PM to PI is much more pronounced in the selected LD spectra compared to the (averaged) powder spectra (see Ref.\onlinecite{rodolakis10}): In the isotropic case the main spectral weight transfer responsible for the A/B ratio change was the one from peak B to higher energies, yet, in the LD data additional selection rules affect also peak A more directly, which has spectral weight \emph{almost} proportional to the amount of $\ket{e_g^{\pi},a_{1g}}$ character in the ground state. This remarkable effect is, of course, due to LD selection rules and it works in a way just the same as the toy--example of the p$_z$ ground state discussed at the beginning of Sec.\ref{sec3}. It can be understood best in the limit of a pure quadrupole transition, where we neglect the mixing of the V 3d--states with the V 4p--states. In this limit the first peak is for our quadrupole transition operators, in fact, displayed \emph{exclusively} in the $\ket{e_g^{\pi},a_{1g}}$ part of the spectrum and completely absent in the $\ket{e_g^{\pi},e_g^{\pi}}$ part. The transitions associated with this peak is the $\ket{e_g^{\pi},a_{1g}}\rightarrow\ket{e_g^{\pi},e_g^{\pi},a_{1g},1s_{\text{core}}}$. Note, if the ground state is $\ket{e_g^{\pi},a_{1g}}$ the $1s$--core electron has to be placed in the free $e_g^{\pi}$ level which can be done. If, however, the ground state is $\ket{e_g^{\pi},e_{g}^\pi}$ the $1s$--core electron has to be placed in the free $a_{1g}$ level -- but this is impossible with the specific transition operators of the displayed spectra, since all the associated matrix elements are integrals over odd functions and thus zero. Hence, we are directly sensitive to the $a_{1g}$ occupation of the ground state. Let us stress that this argument holds strictly only in the abovementioned assumption of no 3d--4p mixing. In the real situation, and also our full calculation, this mixing weakens this effect slightly (the first peak is displayed also in the $\ket{e_g^{\pi},e_g^{\pi}}$ part of the spectrum due to the d--p--hybridization) but nonetheless it is still clearly observable. In conclusion, the experimental LD data of Fig.~\ref{V2O3_mixtable} presents further evidence for the robustness of our ground state probe and their theoretical simulation is consistent with the ground state parameters we found for the isotropic case. 

\section{Summary and outlook}
\label{sec5}
By hands of the well known transition metal oxide V$_2$O$_3$ we have shown that XAS for the transition metal K-edge can provide valuable information on the electronic structure of the $d$-orbitals, even though these are only accessible by quadrupole transitions and hence only form a preedge, which has an order of magnitude less weight than the main dipole-allowed oxygen edge. As a matter of course this low intensity is a disadvantage in comparison to the Vanadium L-edge. However, the hard x-rays of the K-edge can penetrate the diamond anvil cell so that XAS experiments are possible under pressure, e.g., to study the pressure induced Mott-Hubbard transition in Cr-doped  V$_2$O$_3$.

Our theoretical analysis has been based on  LDA and CI calculations. The CI  better accounts for local Coulomb interaction and excitonic effects, both of which are of importance  for the $d$-states of the preedge, whereas LDA is more suitable for  the more extended oxygen states. We elaborated the formalism for the dipole and quadrupole transitions for different polarization directions of the XAS light, and elucidated on the quantum mechanical interferences between both. The dipole transitions are possible since there is a small, non-negligible $4p$ character of the predominantly $3d$ eigenstates in a transition metal oxide. Because of this there are two transitions paths  of comparable magnitude from the core state to the predominantly $3d$ eigenstates: (i) the quadrupole transition to the $3d$ part of the eigenstate and (ii) a dipole transition to the mixed-in oxygen $4p$ part of the eigenstate. Because of quantum mechanics, we have (iii) interferences between the two paths. To the best of our knowledge, these interference processes have not been studied in detail before. Let us note that for a lattice with inversion symmetry these interference terms cancel. However, for V$_2$O$_3$ there is locally no inversion symmetry for each Vanadium atom since there is a neighboring V atom in one direction along the $\mathbf c$ axis but not in the opposite direction. Hence there is a  dipole-quadrupole interference term  in the conductivity tensor (Fig.\ref{V2O3_SigmaArray}) for each Vanadium atom. The whole crystal with 4 Vanadium atoms, on the other hand, is inversion symmetric; and accordingly the interference terms cancel in the average over all four vanadium spectra, i.e., in a XAS experiment. This is different for an x-ray diffraction experiment where, due to the $\mathbf{k}$-dependence, the four atoms contribute with a different phase.

Of particular importance is the study of the polarization dependence of the XAS spectrum, i.e., linear dichroism. This is necessary for the aforementioned interference effects, and also to extract information on the orbital occupations, i.e., in the case of V$_2$O$_3$ on how many $e_g^{\pi}$ and $a_{1g}$ states are occupied and how this changes at the pressure induced Mott-Hubbard transition \cite{rodolakis10}.

With our study, we  demonstrated the usefulness of analyzing preedges in XAS with quadrupole transitions. We have highligted in detail the difficulties which are encountered in such analysis and we showed how to solve them. Hence, we hope to trigger more XAS experiments of this kind as well as related electron energy loss spectroscopy (EELS) experiments for which dichroism measurements recently became possible\cite{SchattschneiderNature}. We also  propose x-ray diffraction experiments for V$_2$O$_3$ to experimentally prove the dipole-quadrupole interference.


\begin{thebibliography}{2pt}
\bibitem{thole92}
B.~T. Thole, P. Carra, F. Sette, and G. van der Laan,
\newblock Phys. Rev. Lett. {\bf 68}, 1943 (1992)

\bibitem{Juhin10}
M. Sikora, A. Juhin, T.-C. Weng, P. Sainctavit, C. Detlefs, F. de Groot and P. Glatzel
\newblock Phys. Rev. Lett. {\bf 105}, 037202 (2010)

\bibitem{chen95}
M. D. N\'u\~nez Regueiro, M.~Altarelli, C.~T. Chen,
\newblock Phys. Rev. B {\bf 51}, 629 (1995)

\bibitem{hansmann08}
P. Hansmann, A. Severing, Z. Hu, M.~W. Haverkort, C.~F. Chang, S. Klein, A. Tanaka, H.~H. Hsieh, H.-J. Lin, C.~T. Chen, B. F\aa{}k, P. Lejay, L.~H. Tjeng,
\newblock Phys. Rev. Lett., \textbf{100}(6), 066405 (2008)

\bibitem{park00}
J.-H. Park, L.~H. Tjeng, A. Tanaka, J.~W. Allen, C.~T. Chen, P. Metcalf, J.~M. Honig, F.~M.~F. de~Groot,  G.~A. Sawatzky,
  \newblock Phys. Rev. B, \textbf{61}(17), 11506 (2000)
  
\bibitem{haverkort05}
M. W. Haverkort, Z. Hu, A. Tanaka, W. Reichelt, S. V. Streltsov, M. A. Korotin, V. I. Anisimov, H. H. Hsieh, H.-J. Lin, C. T. Chen, D. I. Khomskii, and L. H. Tjeng,
  \newblock Phys. Rev. Lett., \textbf{95}, 196404 (2005)  

\bibitem{deGroot_book}
F. de Groot and A. Kotani, \emph{Core Level Spectroscopy of Solids},  Taylor \& Francis CRC press, 2008

\bibitem{fink85}
J.~Fink, T.~M\"uller-Heinzerling, B.~Scheerer, W.~Speier, F.~U. Hillebrecht, J.~C. Fuggle, J.~Zaanen,  G.~A. Sawatzky,
\newblock Phys. Rev. B, \textbf{32}(8), 4899 (1985)

\bibitem[{Thole(1997)}]{thole97}
T.~Thole,
\newblock \emph{Memorial issue},
\newblock Journal of Electron Spectroscopy and Related Phenomena,
  \textbf{86}(1-3), 1  (1997)
  
\bibitem{groot94}
F.~M.~F. de~Groot,
\newblock Journal of Electron Spectroscopy and Related Phenomena,
  \textbf{67}(4), 529  (1994)
  
\bibitem{tanaka94}
A.~Tanaka  T.~Jo,
\newblock Journal of the Physical Society of Japan, \textbf{63}(7), 2788 (1994)

\bibitem{lupinature}
S.~Lupi, L.~Baldassarre, B.~Mansart, A.~Perucchi, A.~Barinov, P.~Dudin, E.~Papalazarou, F.~Rodolakis, J.~-P.~Rueff, J.~-P.~Iti\'e, S.~Ravy, D.~Nicoletti, P.~Postorino, P.~Hansmann, N.~Parragh, A.~Toschi, T.~Saha-Dasgupta, O.K.~Andersen, G.~Sangiovanni, K.~Held, and M.~Marsi,
\newblock Nature Communications {\bf 1}, 105 (2010)

\bibitem{mcwhan69}
D.~B. McWhan, T.~M. Rice, J.~P. Remeika,
\newblock Phys. Rev. Lett., \textbf{23}, 1384 (1969)

\bibitem{robinson75}
W.~R. Robinson,
\newblock Acta Crystallographica Section B, \textbf{31}, 1153 (1975)

\bibitem{toschi10}
A. Toschi, P. Hansmann, G. Sangiovanni1, T. Saha-Dasgupta, O.~K. Andersen, and K. Held
\newblock J. Phys.: Conf. Ser. {\bf 200} 012208 (2010)

\bibitem{castellani78A}
C.~Castellani, C.~R. Natoli, J.~Ranninger,
\newblock Phys. Rev. B, \textbf{18}, 4945 (1978{\natexlab{b}})

\bibitem{castellani78B}
C.~Castellani, C.~R. Natoli,  J.~Ranninger,
\newblock Phys. Rev. B, \textbf{18}, 4967 (1978{\natexlab{a}})

\bibitem{held01B}
K.~Held, G.~Keller, V.~Eyert, D.~Vollhardt, V.~I. Anisimov,
\newblock Phys. Rev. Lett., \textbf{86}, 5345 (2001{\natexlab{a}})

\bibitem{held01}
K.~Held, A.~K. McMahan, R.~T. Scalettar,
\newblock Phys. Rev. Lett., \textbf{87}, 276404 (2001{\natexlab{b}})

\bibitem{tanusri09}
T.~Saha-Dasgupta, O.~K. Andersen, J.~Nuss, A.~I. Poteryaev, A.~Georges, A.~I. Lichtenstein,
\newblock arXiv.org, \textbf{0}, 0907.2841 (2009{\natexlab{a}})

\bibitem{poteryaev07}
A. I. Poteryaev, J. M. Tomczak, S. Biermann, A. Georges, A. I. Lichtenstein, A. N. Rubtsov, T. Saha-Dasgupta, and O. K. Andersen 
\newblock Phys. Rev. B {\bf 76}, 085127 (2007) 

\bibitem{rodolakis10}
F.~Rodolakis, P.~Hansmann, J.-P. Rueff, A.~Toschi, M.~W. Haverkort, G.~Sangiovanni, A.~Tanaka, T.~Saha-Dasgupta, O.~K. Andersen, K.~Held, M.~Sikora, I.~Alliot, J.-P. Itie, F.~Baudelet, P.~Wzietek, P.~Metcalf,   M.~Marsi,
\newblock Phys. Rev. Lett., \textbf{104}, 047401 (2010)

\bibitem{rodolakis11}
F. Rodolakis, J.-P. Rueff, M. Sikora, I. Alliot, J.-P. Itié, F. Baudelet, S. Ravy, P. Wzietek, P. Hansmann, A. Toschi, M.W. Haverkort, G. Sangiovanni, K. Held, P. Metcalf, M. Marsi,
\newblock arXiv:1110.4397v1 (2011)

\bibitem{rodolakis09}
F. Rodolakis, P. Hansmann, J.-P. Rueff, A. Toschi, M.~W. Haverkort, G. Sangiovanni, K. Held, M. Sikora, A. Congeduti, J.-P. Iti\'e, F. Baudelet, P. Metcalf, and M. Marsi 
\newblock J. Phys.: Conf. Ser. {\bf 190} 012092 (2009)

\bibitem{rodolakis09PRL}
F. Rodolakis, B. Mansart, E. Papalazarou, S. Gorovikov, P. Vilmercati, L. Petaccia, A. Goldoni, J. P. Rueff, S. Lupi, P. Metcalf, and M. Marsi
\newblock Phys. Rev. Lett., \textbf{102}, 066805 (2009)


\bibitem{baldassarre08}
L. Baldassarre, A. Perucchi, D. Nicoletti, A. Toschi, G. Sangiovanni, K. Held, M. Capone, M. Ortolani, L. Malavasi, M. Marsi, P. Metcalf, P. Postorino, and S. Lupi 
\newblock Phys. Rev. B {\bf 77}, 113107 (2008)

\bibitem{stohr_book}
J. St\"ohr, \emph{NEXAFS Spectroscopy}, Springer (1996)



\bibitem{moon70}
R.~M. Moon,
\newblock Phys. Rev. Lett., \textbf{25}(8), 527 (1970)

\bibitem{paolasini99}
L.~Paolasini, C.~Vettier, F.~de~Bergevin, F.~Yakhou, D.~Mannix, A.~Stunault,  W.~Neubeck, M.~Altarelli, M.~Fabrizio, P.~A. Metcalf,  J.~M. Honig,
\newblock Phys. Rev. Lett., \textbf{82}, 4719 (1999)

\bibitem{dimateo02}
S.~Di~Matteo, N.~B. Perkins,  C.~R. Natoli,
\newblock Phys. Rev. B, \textbf{65}, 054413 (2002)

\bibitem{loudon83}
R.~Loudon,
\newblock \emph{The Quantum Theory of Light},
\newblock (Oxford: Clarendon) (1983)

\bibitem{mercouris97}
T.~Mercouris, Y.~Komninos, S.~Dionissopoulou,  C.~A. Nicolaides,
\newblock Journal of Physics B: Atomic, Molecular and Optical Physics,
  \textbf{30}(9), 2133 (1997)

\bibitem{Tsvetkov04}
A. A. Tsvetkov, F. P. Mena, P. H. M. van Loosdrecht, D. van der Marel, Y. Ren, A. A. Nugroho, A. A. Menovsky, I. S. Elfimov, and G. A. Sawatzky 
\newblock Phys. Rev. B {\bf 69}, 075110 (2004)

\bibitem{SchattschneiderNature}
P. Schattschneider, S. Rubino, C. H\'ebert, J. Rusz, J. Kune, P. Nov\'ak, E. Carlino, M. Fabrizioli, G. Panaccione, and G. Rossi
\newblock Nature {\bf 441}, 486 (2006)

  \end{thebibliography}
\end{document}